\documentclass[prc,reprint,twocolumn,amsmath,amssymb,groupedaddress]{revtex4}%-2}

\usepackage{graphicx}% Include figure files
\usepackage{dcolumn}% Align table columns on decimal point
\usepackage{bm}% bold math
\usepackage{tabularx}
\usepackage{multirow}
\usepackage{hyperref}

\begin{document}
%\begin{CJK*}{GBK}{song}
%\preprint{APS/123-QED}

\title{Relativistic Mean-Field Approach in Nuclear Systems}% Force line breaks with \\

\author{Xiaodong Sun}\affiliation{China Institute of Atomic Energy, P.O. Box 275(41), Beijing 102413, China}%\affiliation{School of Nuclear Science and Technology, Lanzhou University, Lanzhou 730000, China} 
\author{Ruirui Xu}\email{xuruirui@ciae.ac.cn}\affiliation{China Institute of Atomic Energy, P.O. Box 275(41), Beijing 102413, China}
\author{Yuan Tian}\affiliation{China Institute of Atomic Energy, P.O. Box 275(41), Beijing 102413, China}
\author{Hongfei Zhang}\email{zhanghongfei@lzu.edu.cn}\affiliation{School of Nuclear Science and Technology, Lanzhou University, Lanzhou 730000, China} 
\author{Zhongyu Ma}\affiliation{China Institute of Atomic Energy, P.O. Box 275(41), Beijing 102413, China}
\author{Zhi Zhang}\affiliation{China Institute of Atomic Energy, P.O. Box 275(41), Beijing 102413, China}
\author{Zhigang Ge}\affiliation{China Institute of Atomic Energy, P.O. Box 275(41), Beijing 102413, China}
\author{E. N. E. van Dalen}\affiliation{Institut f\"{u}r Theoretische Physik, Universit$\ddot{a}$t T$\ddot{u}$bingen, Auf der Morgenstelle 14, D-72076 T$\ddot{u}$bingen, Germany}
\author{H. M\"{u}ther}\email{herbert.muether@uni-tuebingen.de}\affiliation{Institut f\"{u}r Theoretische Physik, Universit$\ddot{a}$t T$\ddot{u}$bingen, Auf der Morgenstelle 14, D-72076 T$\ddot{u}$bingen, Germany}

\date{\today}% It is always \today, today,

\begin{abstract}
A new scheme to study the properties of finite nuclei is proposed based on the Dirac-Brueckner-Hartree-Fock (DBHF) approach starting from a bare nucleon-nucleon interaction. The relativistic structure of the nucleon self-energies in nuclear matter depending on density, momentum and isospin asymmetry are determined through a subtracted T-matrix technique and parameterized, which makes them easily accessible for general use. The scalar and vector potentials of a single particle in nuclei are generated via a local density approximation (LDA). The surface effect of finite nuclei can be taken into account by an improved LDA (ILDA), which has successfully been applied in microscopic derivations of the optical model potential for nucleon-nucleus scattering. The bulk properties of  nuclei can be determined in a self-consistent scheme for nuclei all over the nuclear mass table. Calculated binding energies agree very well with the empirical data, while the predicted values for radii and spin-orbit splitting of single-particle eneries are about 10 \% smaller than the experimental data. Basic features of more sophisticated DBHF calculations for finite nuclei are reproduced. 
\end{abstract}

%\pacs{24.10.Ht;24.10.Cn;24.10.Jv;21.65.Cd}% PACS, the Physics and Astronomy
\maketitle

\section{Introduction}
One of the central challenges of theoretical nuclear structure physics is to derive the bulk properties of nuclear systems from a realistic nucleon-nucleon (NN) interaction, i.e. a model of the NN interaction, which is adjusted to the data of NN scattering. Various models for such realistic NN have been developed over the years including local potentials as the Argonne potential\cite{argonne}, various approaches based on the meson-exchange or One-Boson-Exchange (OBE) picture\cite{erkelenz,mhelster,Machxxx} or those based on chiral effective field theory\cite{kaiserbw,epelbaum,machchi}. 

In a first step one tries to evaluate the energy per nucleon in infinite nuclear matter as a function of density. For isospin symmetric nuclear matter one would like to reproduce the so-called saturation point, i.e. the empirical values for the minimum of the energy as a function of density curve, which, according to volume term of the Bethe-Weizs\"acker mass formula\cite{we35,bethe36}, should occur at $-16$ MeV per nucleon at a nuclear density around 0.17 fm$^{-3}$.

One finds that all non-relativistic many-body calculations of nuclear matter using a realistic two-nucleon interaction yield saturation points for nuclear matter which are all positioned in a band in the energy {\it vs.} density plane, the so-called Coester band\cite{coester1,coester2}, which does not meet the empirical point. Three-nucleon (3N) interactions have to be introduced in order to reproduce the bulk properties of infinite nuclear matter\cite{3nuc1,3nuc2,3nuc3}.

On the other hand, relativistic calculations have been successful in describing the empirical saturation point of symmetric nuclear matter from realistic meson exchange interactions without introducing phenomenological many-nucleon forces\cite{mike,horow84,rolf84,malf87,weigel88,lenske,erik04}. The mechanism, which leads to the empirical saturation point in these Dirac-Brueckner-Hartree-Fock (DBHF) calculations can be described in terms of the phenomenological Walecka model\cite{walecka}. The exchange of a scalar meson, $\sigma$, leads in the mean field approach to a strong attractive component in the nuclear self-energy, which transforms like a scalar under a Lorentz transformation. In the nuclear binding this attractive component is compensated to a large extent by a repulsive component, which transforms like the time-like component of a Lorentz vector and originates from the interaction with a vector meson, $\omega$. Introducing this self-energy into the Dirac equation for the nucleon, one obtains Dirac spinors in the nuclear medium, with an enhanced small component as compared to the corresponding Dirac spinors for the nucleon in the vacuum. Including this density dependence of the nucleon Dirac spinors in evaluating the NN interaction of a realistic OBEP one gets a small repulsive effect increasing with density, which is sufficient to shift the calculated saturation point to the position derived from experimental data. Attempts have been made to simulate this effect in terms of a density-dependence of the NN force or an explicit (3N) interaction\cite{francesca}.

Studies of infinite nuclear matter are only one step towards a microscopic understanding of nuclear structure. Many attempts have been made to apply the DBHF approximation also to the description of finite nuclei. This is more complicate as it requires an evaluation of the Dirac spinors for finite nuclei and a solution of the  two-nucleon equation in the nuclear medium in terms of these spinors. This has been achieved only recently by Shen $et$ $al$. \cite{shen1,shen2} 

Because of the complications mentioned above a lot of effort has been made to take advantage of the DBHF results obtained in nuclear matter and use various kinds of local-density approximations in the evaluation of finite nuclei\cite{brockm,boersma,fritz1,fritz2,toki,zyma}. As an example we mention various attempts to determine an effective meson theory with coupling constants, depending on the nuclear density. These effective coupling constants are adjusted to reproduce the results of DBHF calculations of nuclear matter and have been used for mean-field or Dirac Hartree Fock calculations of finite nuclei\cite{fritz2,zyma}.

A different kind of local density approach has been developed  to derive the optical model potential for nucleon-nucleus scattering from DBHF calculations of isospin symmetric and asymmetric nuclear matter\cite{ruirui1,ruirui2}. The basis of this approach is the nucleon self-energy depending on the nucleon energy. For positive energies, which means nucleon energies above the corresponding Fermi energy, the DBHF approach leads to complex values  for the scalar, time- and space-like vector component of the self-energy. These components are then used to determine the optical model potential to be used in a Schroedinger equation for elastic nucleon-nucleus scattering assuming a specific density profile for the proton and neutron distributions of the target nucleus under consideration. 

This scheme is an extension of the microscopic optical model developed by Jeukenne $et$ $al$. \cite{Jeu77} which uses the non-relativistic Brueckner-Hartree-Fock (BHF) approximation for the nucleon self-energy. It is one advantage of the relativistic scheme that it also yields a spin-orbit term for nucleon-nucleus scattering. The results of the relativistic approach have been very promising. The empirical data for elastic proton-nucleus and neutron-nucleus scattering can be reproduced with good accuracy for a broad mass range of target nuclei and a large energy region of scattered nucleons\cite{ruirui2}. Therefore the resulting CTOM (China Nuclear Data Center and T\"ubingen University Optical Model) has been made available in terms of a tool, which yields the optical model potential for density distributions of the target nucleus either supplied by the user or suggested by the tool\cite{ctomtool}.

Motivated by this success of CTOM we would like to explore in the work presented here, whether the same local density approximation used to derive the optical model potential of CTOM can also be used to determine the bulk properties of nuclei in a self-consistent way. For that purpose we will briefly review the evaluation of the nucleon self-energy in the DBHF approach and present a simple parameterization of components of this self-energy for bound nucleons in Section II of this paper. The use of the local density approximation and the self-consistent solution of the Dirac equation is discussed in Section III. Section III also contains the presentation and discussion of results for spherical nuclei from $^{16}$O to $^{208}$Pb. Special attention will be paid to the effects of the nuclear surface, which are not contained in the density dependence of infinite matter results, but introduced in terms of a finite range correction of an Improved Local Density Approximation (ILDA) suggested by Jeukenne $et$ $al$. \cite{Jeu77}. Finally, the overall discussion is summarized in Section IV.

\section{SELF-ENERGY IN NUCLEAR MATTER}

The relativistic structure of the nucleon self-energy derived in infinite nuclear matter in the DBHF approximation can be written in the form
\begin{eqnarray}
     \Sigma^\tau(E,k,\rho,\beta) & =  &\Sigma^\tau_\mathrm{S}(k,E,\rho,\beta) - \gamma_0\Sigma_0^\tau(k,E,\rho,\beta)\nonumber \\
&&      + \bm{\gamma}\cdot
     \textbf{k}\Sigma_\mathrm{V}^\tau(k,E,\rho,\beta)~. \label{eq5}
\end{eqnarray}
In this equation, $\Sigma_\mathrm{S}$ is the scalar part of self-energy, $\Sigma_0$ and
$\Sigma_\mathrm{V}$ denote the time-like and space-like terms of the vector part, respectively. The superscript $\tau$ is used
to identify the isospin, since protons and neutrons should be distinguished in
isospin asymmetric nuclear matter. Note that these components of the
self-energy are functions of the nucleon momentum ($k$), the single-particle energy $E$, the density $\rho$  and asymmetry parameter 
\begin{equation}
\beta = \frac{\rho_n - \rho_p}{\rho}\,,\label{defbeta}
\end{equation}
where $\rho_n$, $\rho_p$ and $\rho$
denote the neutron, proton and total densities in nuclear matter, respectively. The energy variable $E$ is normalized
in such a way that $E = 0$ corresponds to the Fermi energy at density $\rho$ and asymmetry
$\beta$ under consideration. The analysis of Ruirui Xu $et$ $al$. \cite{ruirui2} is based on the Bonn B potential\cite{Machxxx} and
employs the subtracted T-matrix representation as described in Ref. \cite{subtrt} to extract the Dirac components of the self-energy. In the present investigation we have chosen the Bonn A potential, which yields better results for the saturation point of isospin symmetric nuclear matter. We will only consider on-shell results for the self-energy, which means that the energy variable $E$ corresponds to the single-particle energy resulting from the BHF self-consistency requirement. Therefore the redundant momentum variable will be dropped in the text below.

\begin{figure}[htb]
  \centering
  \includegraphics[width=0.48\textwidth]{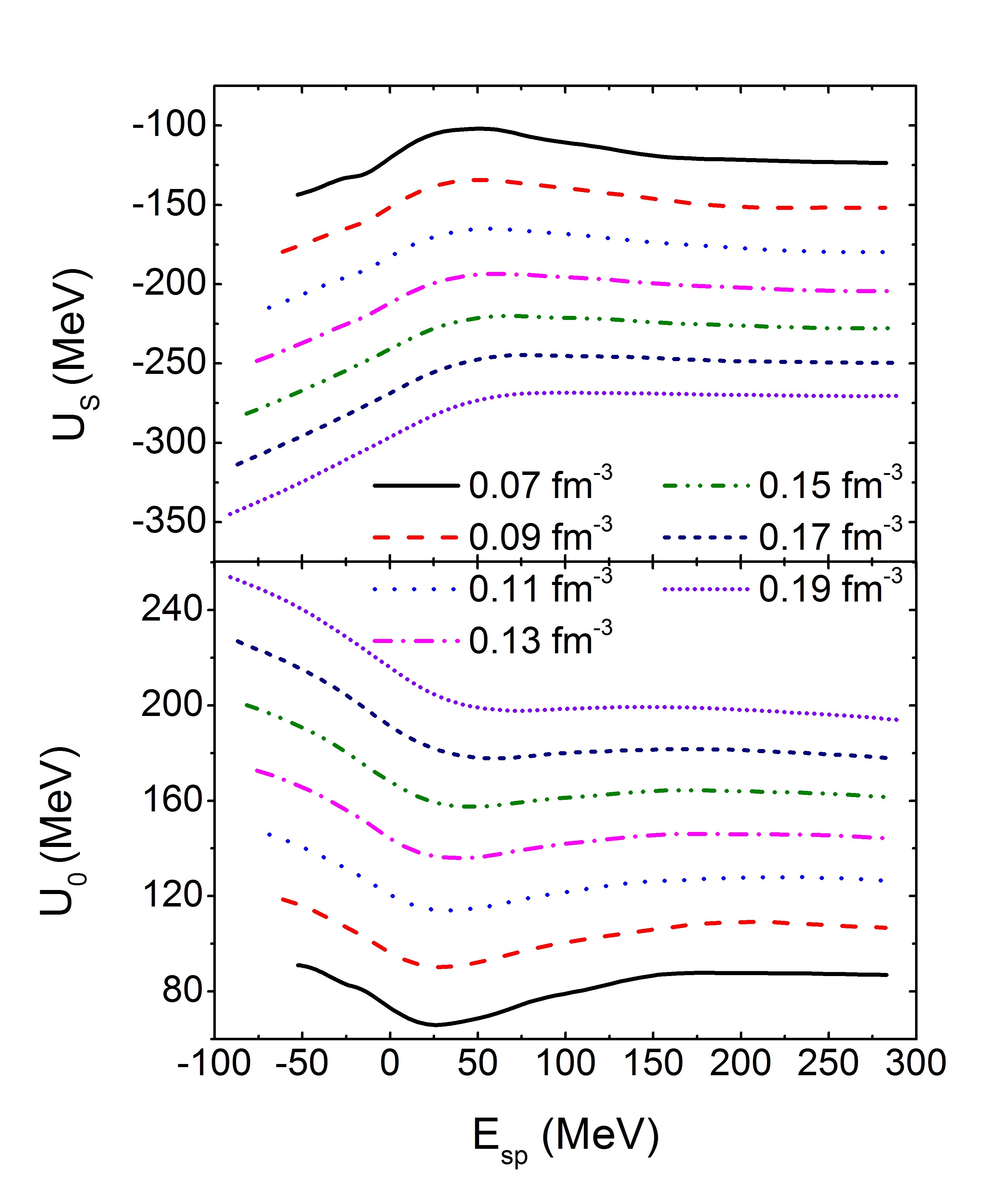}
  \caption{Dirac scalar and vector potentials as a function of nucleon single particle energy $E$ at various densities for symmetric nuclear matter resulting from DBHF calculations\cite{subtrt} using the potential Bonn A \cite{Machxxx}. }\label{Unmjpg}
\end{figure}

From the Dirac components of the self-energy defined in eq.(\ref{eq5}) one may define the scalar and vector potentials
\begin{eqnarray}
U^\tau_\mathrm{S}=\frac{\Sigma^\tau_\mathrm{S}-\Sigma^\tau_\mathrm{V} M}{1+\Sigma^\tau_\mathrm{V}},\nonumber\\
U^\tau_0=\frac{-\Sigma^\tau_0+\varepsilon\Sigma^\tau_\mathrm{V}}{1+\Sigma^\tau_\mathrm{V}},\label{eq6}
\end{eqnarray}
with $M$ the mass of the nucleon and the energy variable $\varepsilon$, which is related to the single-particle energy E by $\varepsilon = E + M$. In terms of these scalar and vector potentials, the Dirac equation for a nucleon in the nuclear mean field can be written
\begin{eqnarray}
      \Bigl[\vec{\alpha}\cdot\vec{p}+\gamma_0(\textit{M}+U^\tau_\mathrm{S})+U^\tau_0 \Bigr]
      \Psi^\tau = \varepsilon\Psi^\tau~ . \label{eq10}
\end{eqnarray}
Results for the real part of the Dirac potentials $U_\mathrm{S}$ and $U_0$ are presented in the upper and lower panel of Fig. \ref{Unmjpg} as a function of the single-particle energy $E$ for various densities. Note that in the case of isospin symmetric nuclear matter, which is considered in this figure, the potentials are identical for protons and neutrons, so that we can drop the superscript $\tau$. 

It is worth noting that the energy dependence is quite different for positive energies $E>0$, which are relevant for the optical model potential, and negative energies, which enter the evaluation of bound states. For negative energies, the nuclear potentials are only real and we obtain a rather simple parameterization of these potentials in the form
\begin{eqnarray}\label{equsym}
U(E,\rho,\beta=0)_{S(0)} &=&\bigl(b_{11} E+b_{12} \bigr) \rho^{2/3}\nonumber\\
&&+\bigl(b_{21} E+b_{22}\bigr)\rho~,
\end{eqnarray}
The parameter of this fit are given in table \ref{tabus0}.
\begin{table}
  \centering
  \caption{Parameters of equation (\ref{equsym}) defining the Dirac potentials $U_{S(0)}$ in symmetric nuclear matter depending on energy and density. The energies are defined units of MeV and the nuclear density, $\rho$, should be defined in fm$^{-3}$.} 
  \label{tabus0}
  \setlength{\tabcolsep}{4pt}
  \begin{tabular}{ccccc}
    \hline
    \hline
&&&&\\
    &$b_{11}$ [fm$^2$]&$b_{12}$ [MeV$\cdot$fm$^2$]&$b_{21}$ [fm$^3$]&$b_{22}$[MeV$\cdot$fm$^3$]\\
&&&&\\
    \hline
&&&&\\
	$U_S$&	2.8249&	-345.68& -2.0445& -979.33  \\
	$U_0$&	-2.931&	-73.42&	2.5649&	1264.5 \\
&&&&\\
    \hline
    \hline
  \end{tabular}
\end{table}
For a better description of the high-density part of the scalar potential $U_S$, the expression (\ref{equsym}) is supplemented by
\begin{equation}
\Delta U_S =756.92\times(\rho-0.16)^{2.01665},~~~~~\rho>0.16 ~\text{fm}^{-3},\label{equsyma}
\end{equation}
which should be added to eq.(\ref{equsym}) for densities $\rho > 0.16 ~\text{fm}^{-3}$.

\begin{figure}[htb]
  \centering
  \includegraphics[width=0.48\textwidth]{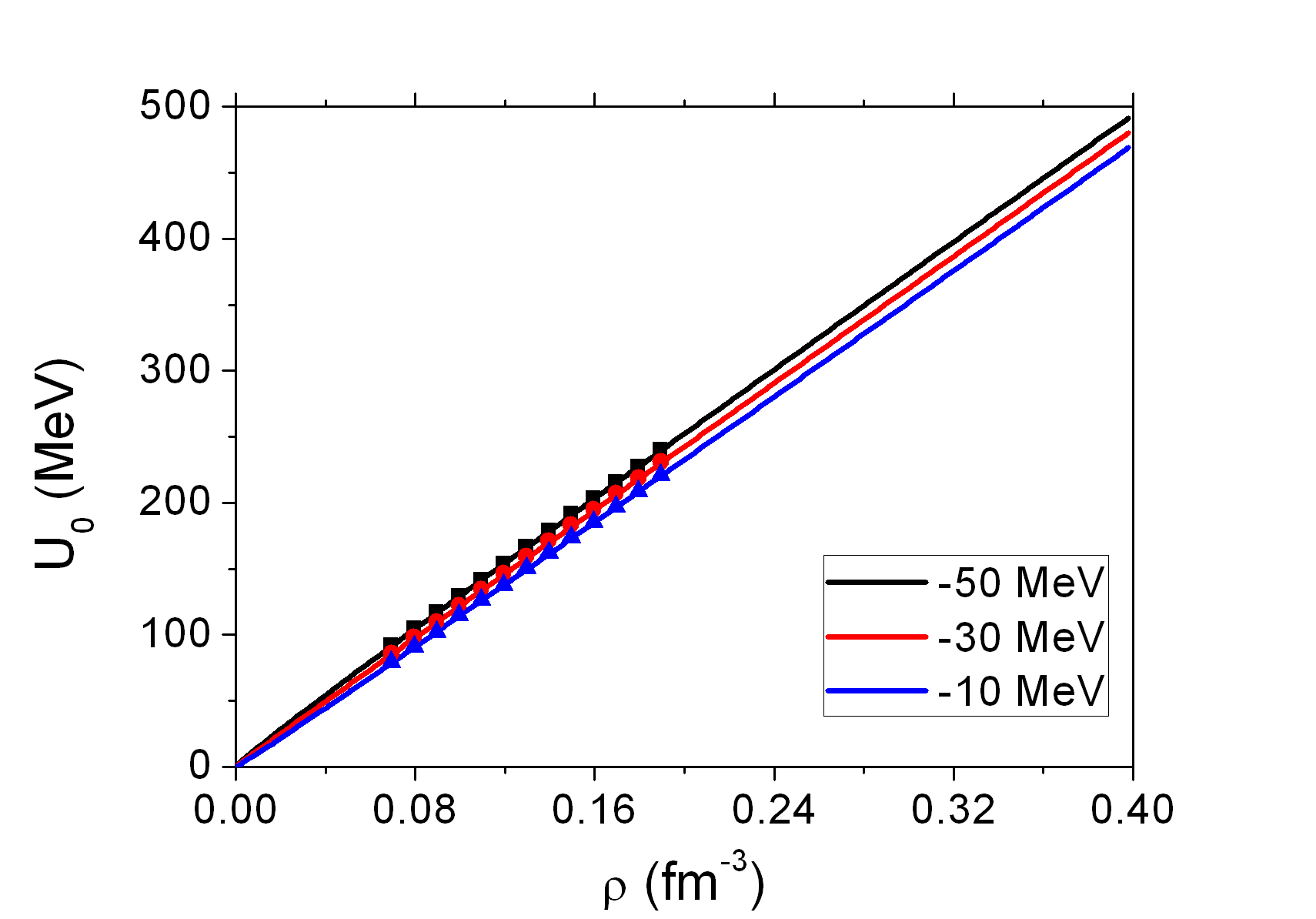}
  \includegraphics[width=0.48\textwidth]{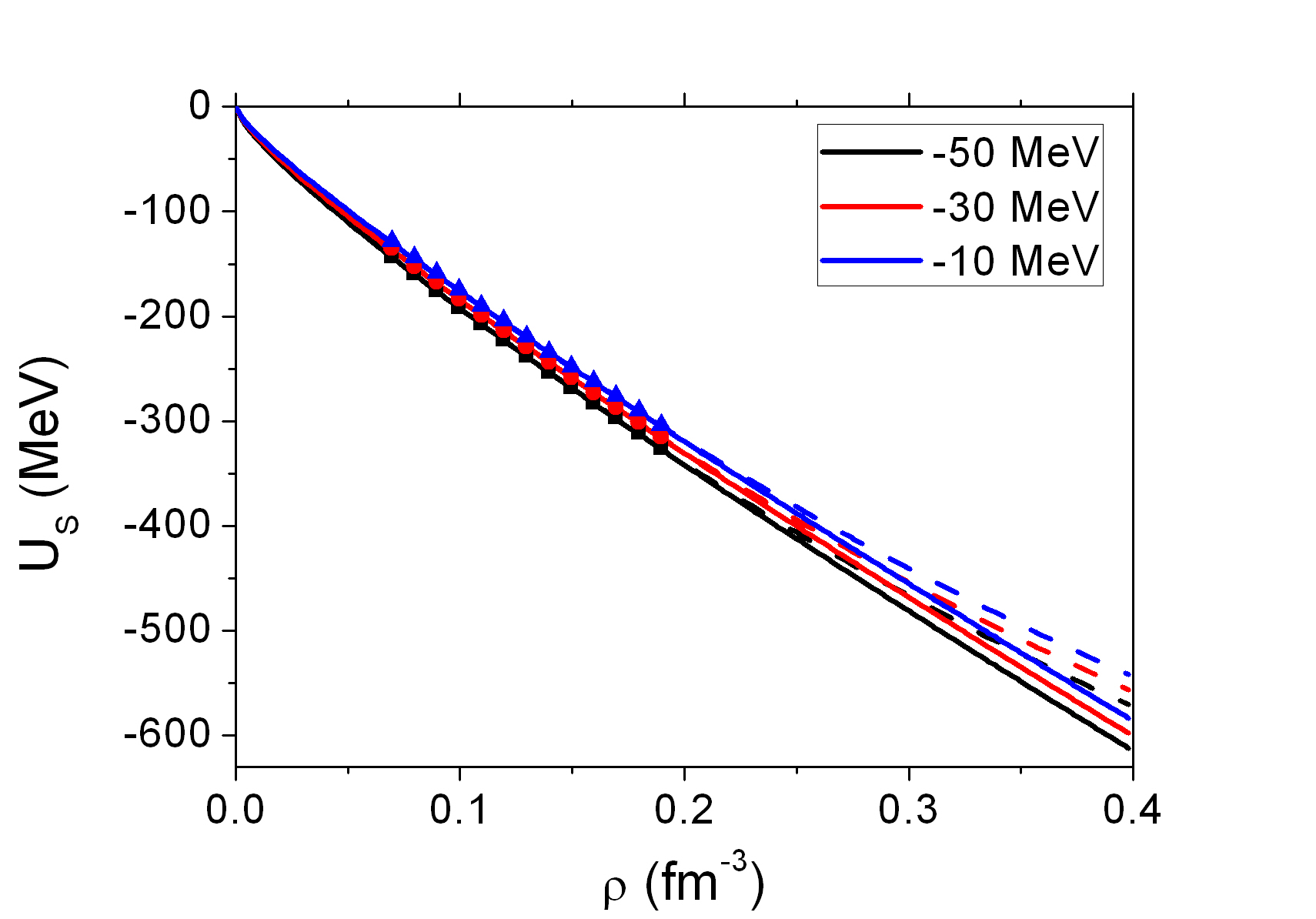}
  \caption{Parameterized curves for the density and energy dependence of the Dirac potentials $U_{S(0)}$ fitting the calculated data points resulting from DBHF approach. The dashed lines at high-densities for the scalar potentials indicate the high-density correction of eq.(\ref{equsyma}).}\label{us0jpg}
\end{figure}

Results of this parameterization are displayed in Fig. \ref{us0jpg} for three different energies as a function of density and compared to corresponding data points of the underlying DBHF calculations. It should be noted that DBHF calculations at small densities ($\rho < 0.07$ fm$^{-3}$) do not provide very reliable data. This is due to the fact that the solution of the Bethe-Goldstone equation suffers from the appearance of quasi-bound two-nucleon states as has also been observed in non-relativistic calculations\cite{pair1}. 

\begin{figure}[htb]
  \centering
  \includegraphics[width=0.48\textwidth]{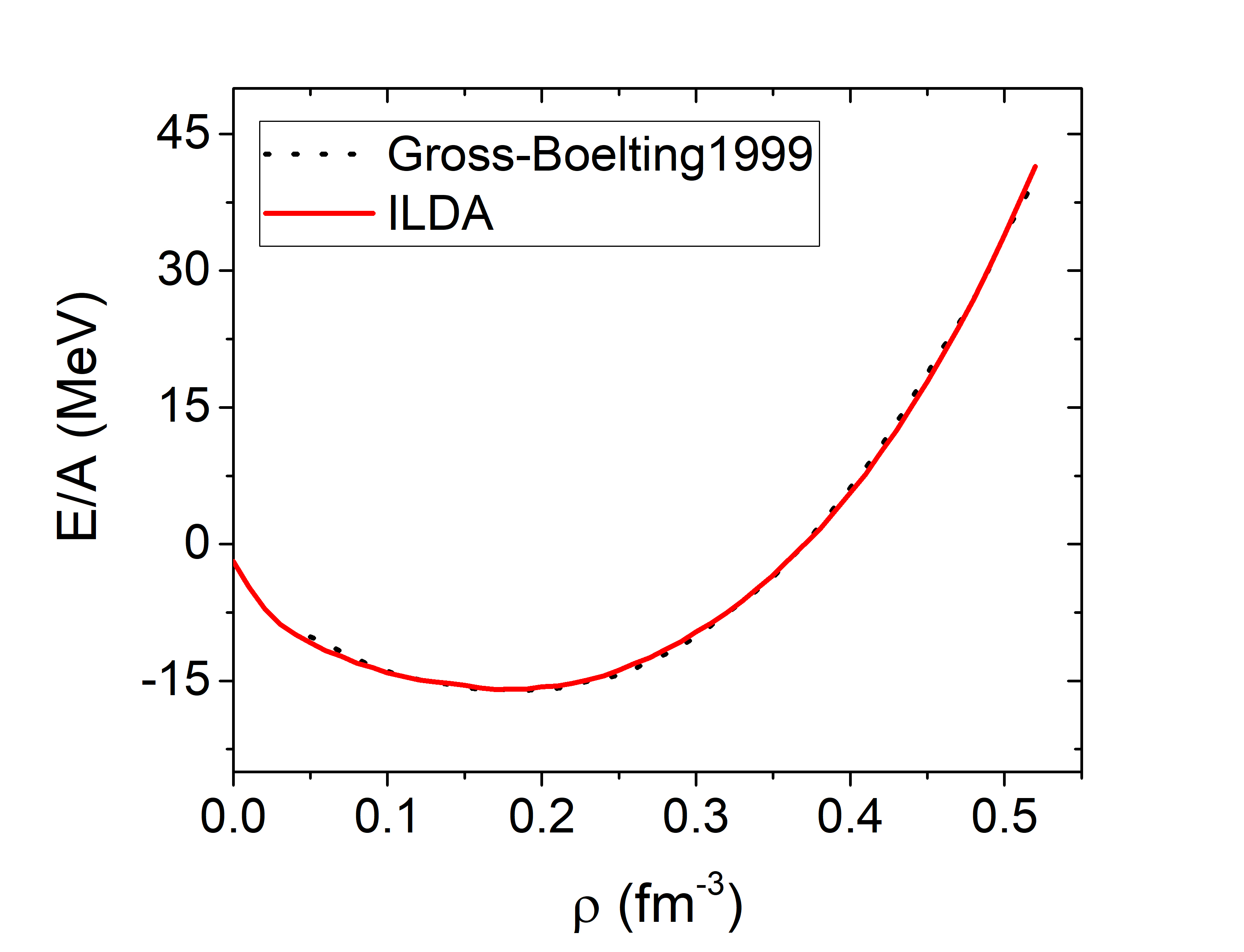}
  \caption{Comparison of binding energy per nucleon in nuclear matter using the parameterization of the Dirac potentials presented in this work and the original DBHF results of Ref. \cite{Gross99}.}\label{eos}
\end{figure}

In order to test this parameterization of the Dirac potentials for symmetric nuclear matter, we have used it to evaluate the self-consistent single-particle energies and the energy per nucleon of nuclear matter. The results of this test are displayed in Fig.~\ref{eos} and compared to the corresponding results of DBHF calculations of Gross-Boelting $et$ $al$. \cite{Gross99}. The agreement is remarkable. Note that the total energy is rather sensitive  as it results from a partial cancellation of the contributions originating from $U_S$ and $U_0$. 

In a next step we now generalize the parameterization for symmetric nuclear to the case of asymmetric matter. In the case of asymmetric nuclear matter the Dirac potentials are in general different for protons and neutrons, identified by the superscript $\tau$. It urns out that these potentials depend to very good approximation linearly on the asymmetry parameter $\beta$ defined in eq.(\ref{defbeta}). Therefore we can parameterize them in the form
\begin{eqnarray}\label{isospineq}
U^\tau_{S(0)} & = & U(\beta=0)_{S(0)}\times\\&&\Bigl\{1+\bigl[(c^\tau_1 E+c^\tau_2)\rho+(c^\tau_3 E+c^\tau_4)\bigr]\beta\Bigr\}. \nonumber
\end{eqnarray}
In this equation $U_{S(0)}$ refers to the corresponding Dirac potential for symmetric nuclear matter. The parameters $c^\tau_i$ are listed in table \ref{tabbeta}.
\begin{table*}
  \centering
  \caption{Parameters for isospin dependence of the Dirac potentials $U_{S(0)}$}
  \label{tabbeta}
  \setlength{\tabcolsep}{6pt}
  \begin{tabular}{cccccc}
    \hline
    \hline
&&&&&\\
    &&$c_1^\tau$ (fm$^3\cdot$MeV$^{-1}$)&$c_2^\tau$ (fm$^3$)&$c_3^\tau$ (MeV$^{-1}$)&$c_4^\tau$\\
&&&&&\\
    \hline
&&&&&\\
\multirow{2}{*}{$U_S$}&proton&0.0166&1.035&-0.0031&-0.5196\\
&neutron&-0.0003&-0.5487&-0.0002&0.5133\\
&&&&&\\
\hline
&&&&&\\
\multirow{2}{*}{$U_0$}&proton&0.0147&2.771&-0.0036&-1.0917\\
&neutron&-0.0091&-2.641&0.0013&1.1786\\
&&&&&\\
    \hline
    \hline
  \end{tabular}
\end{table*}

\begin{figure*}[htb]
  \centering
  \includegraphics[width=0.7\textwidth]{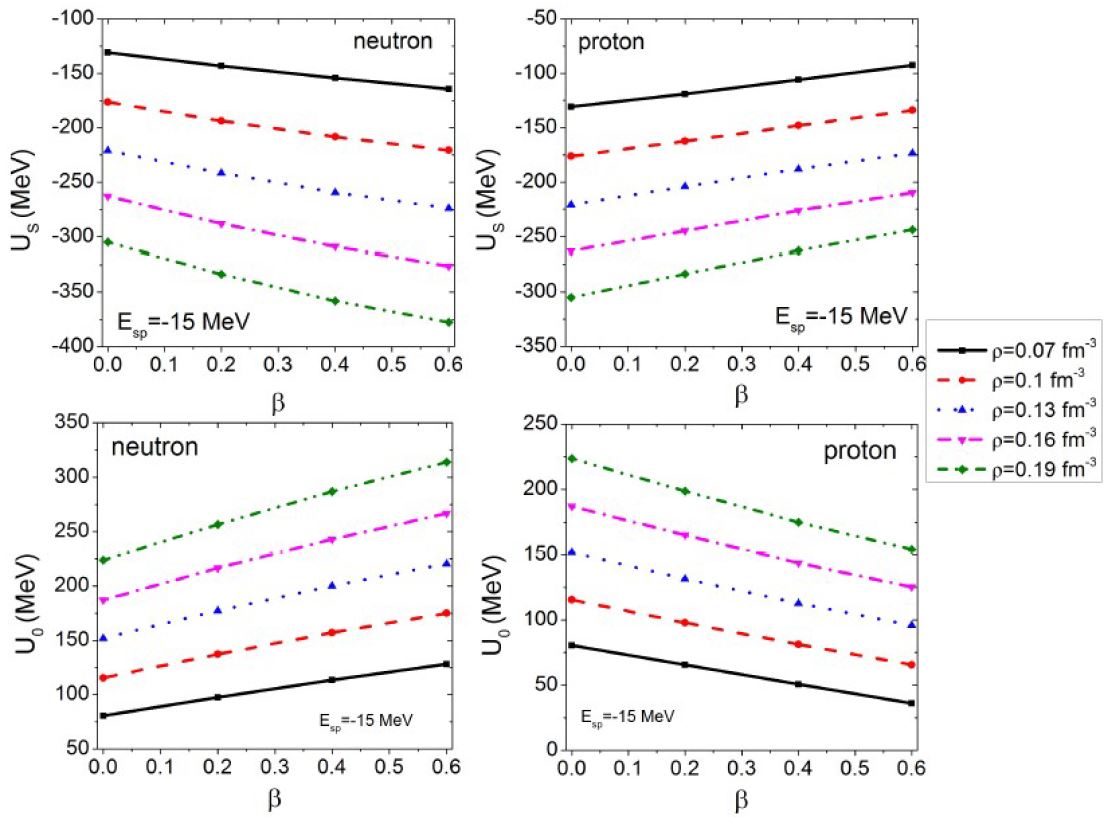}
  \caption{Scalar ($U_S$) and vector potential $U_0$ for neutrons (left column) and protons (right column) as a function of asymmetry $\beta$. Results are presented for a given energy $E = -15$ MeV and various values of the density $\rho$.}\label{uisospin}
\end{figure*}

Results for the Dirac potentials $U_S$ and $U_0$ for protons and neutrons as a function of asymmetry $\beta$ are displayed in Fig. \ref{uisospin} for a given energy ($E = -15$ MeV) at various densities. For positive values of $\beta$ the scalar potential for neutrons is decreasing as a function of $\beta$ whereas an increase is observed for protons. As the Dirac mass for the nucleons is given by
\begin{equation}
M^*_{\tau, D} = M + U^\tau_S ,
\end{equation}
this implies that the effective Dirac mass for neutrons is smaller than the one for protons ($\beta > 0$). This opposite to the dependence of the effective mass deduced from the momentum dependence of the single-particle energy in asymmetric nuclear matter\cite{dalen2011}.

\section{RESULTS FOR FINITE NUCLEI}
\subsection{LOCAL DENSITY APPROXIMATION}
With the parameterization of the Dirac potentials the Dirac equation can be solved self-consistently. For a given nucleus we first assume a nuclear density, $\rho(\bm r)$, and isospin asymmetry distribution, $\beta(\bm r)$, as well as the energies for each single particle state $i$, $E_i$, as initial values for an iteration. With these assumptions we can determine the nucleon scalar and vector potentials using the Local Density Approximation (LDA) and evaluate
\begin{equation}
U_{i,S(0)} (\bm r) = U_{S(0)}^{\tau_i} \bigl( E_i,\rho(\bm r),\beta(\bm r)\bigr) \label{deflda},
\end{equation}
where $\tau_i$ denotes the isospin of the single-particle state $i$. Using these local Dirac potentials, supplemented by the corresponding Coulomb term in the case of proton states, one can solve the Dirac equation for the single-particle states, eq.(\ref{eq10}), and determine a new set of single-particle energies $E_i$ and Dirac spinors $\Psi_i(\bm r)$. These Dirac spinors also determine an update for the single-particle densities, $\rho_i(\bm r)$ as well as the corresponding isospin distributions $\beta_{i}(\bm r) $. Summing these densities over all states $i$ below the Fermi surface $F$ this also yields new results for the density profiles $\rho( \bm r)$, which enter the LDA definition in eq.(\ref{deflda}). This procedure can be iterated until a self-consistent solution is obtained. Note that in this paper we are restricting our studies to the case of spherical nuclei with closed shells for protons and neutrons.

Finally, one can also determine the total energy of the nucleus using
\begin{eqnarray}\label{eqEnergy}
E&=&\sum_{i<F} E_{i}\nonumber\\&&-\frac{1}{2}\sum_{i<F}\int [U_{i,S}(\bm{r})\rho_{S,i}(\bm{r})+U_{i,0}(\bm{r})\rho_i(\bm{r})]d^3\bm{r}\nonumber\\
&&-\frac12\int A_0(\bm{r})\rho_C(\bm{r})d^3\bm{r}+E_{CM},
\end{eqnarray}
where $A_0$ and $\rho_C$ denote Coulomb potential and charge density, respectively. The center of mass correction energy is taken as $E_{CM}=-\frac{3}{4}41A^{-1/3}$. 

\begin{table*}
  \centering
  \caption{Calculated energy per nucleon and radii of charge distribution, $R_C$, of 7 spherical nuclei in LDA and ILDA compared with the experimental data. The last column presents the value of Gaussian width parameter $t$ to be used in eq.(\ref{eqt}).}
  \label{tabfn}
  \setlength{\tabcolsep}{6pt}
  \begin{tabular}{cccccccc}
    \hline
    \hline
&&&&&&&\\
{Nuclei}	&$R_C^\text{expt}$ &	$E^\text{expt}/A$ &	$R_C^\text{LDA}$ &	$E^\text{LDA}/A$ &	$R_C^\text{ILDA}$ &	$E^\text{ILDA}/A$ &$t$	\\
	& [fm]&[MeV]&[fm]&[MeV]&[fm]&[MeV]&[fm]\\
&&&&&&&\\
    \hline
&&&&&&&\\
$^{16}$O	&2.70&	-7.98&	2.34&	-12.25&	2.46&	-7.97&1.00  \\
$^{40}$Ca	&3.48&	-8.55&	3.04&	-11.28&	3.10&	-8.57&0.93\\
$^{48}$Ca	&3.50&	-8.67&	3.17&	-10.70&	3.16&	-8.67&0.84 \\
$^{90}$Zr	&4.30&	-8.71&	3.87&	-10.02&	3.86&	-8.72&0.74 \\
$^{116}$Sn	&4.62&	-8.52&	4.22&	-9.46 &	4.19&	-8.52&0.67 \\
$^{132}$Sn	&4.79&	-8.35&	4.42&	-9.18 &	4.35&	-8.36&0.66 \\
$^{208}$Pb	&5.50&	-7.87&	5.13&	-8.35 &	5.08&	-7.88&0.52 \\
&&&&&&&\\
    \hline
    \hline
  \end{tabular}
\end{table*}
Results for the energy per nucleon $E^{LDA}/A$ and the radius of the charge distribution, $R_C^{LDA}$, are presented in table \ref{tabfn} for 7 spherical nuclei ranging from $^{16}$O to $^{208}$Pb. The radius of the charge distribution has been evaluated from the radius of the proton distribution $R_{\rm pro}$ corrected for the finite size of the proton with the relationship $R^2_C=R^2_{\rm pro}+0.64$, with all radii given in units fm. 

It is true of course that the LDA considered so far includes, what we would like to call for the discussion of this section a ``surface density'' effect. This means it takes into account that the binding energy per nucleon of infinite matter decreases with density for densities below the saturation density. As the average density of finite nuclei is smaller than the saturation density, this leads to binding energies per nucleon for the LDA of finite nuclei, which are smaller than the saturation energy calculated for infinite nuclear matter. 

Comparing these results of the simple LDA with the experimental data, which are also listed in table \ref{tabfn} one finds that the LDA approach overestimates the binding energy per nucleon to a large extent. This is true in particular for light nuclei as $^{16}$O and the isotopes of Calcium. This dependence of the mismatch on the nucleon number indicates that the LDA considered so far does not account for the surface effects in an appropriate way.

\subsection{FINITE RANGE EFFECT}
We took into account the finite range of the NN interaction, an effect which seems to be missing in the LDA discussed so far. Therefore we follow the idea of Jeukenne $et$ $al$. \cite{Jeu77}, which has also been used by others \cite{ruirui2,jamin82} in the derivation of a microscopic approach for the optical model potential of nucleon-nucleus scattering, and account for finite-range effects by using the Improved Local Density Approximation (ILDA) defining
\begin{equation}\label{eqt}
U_{S(0)}^{\rm ILDA}({\bm r})=\frac{1}{(t\sqrt{\pi})^3}\int U_{S(0)}(\bm{r'})\exp\left[-({\bm r}-{\bm r'})^2/t^2\right]{\rm{d}}^3 {\bm r'},
\end{equation}
where $U_{S(0)}(\bm{r'})$ refers to the LDA approximation of eq. (\ref{deflda}) and $t$ denoting the Gaussian width is a parameter to be adjusted. The denominator guarantees the volume integral unchanged. The larger $t$, the smoother the potentials varying with the radius $r$, and the stronger the potentials at surface region. In our study we fit this parameter $t$, to reproduce the binding energy of finite nuclei. Results for the calculated energy per nucleon and the radius of the charge distribution as well as the value of the width parameter $t$ are also listed in table \ref{tabfn}.

Using values for $t$ which are smoothly varying with the mass number $A$ and in reasonable agreement with those derived from fits of the optical model data we are able to reproduce the binding energy per nucleon. Also the calculated radii of the charge distribution are improved as compared to the LDA approach discussed above, but still almost 10 \% smaller compared to the experimental data.

\begin{figure*}[htb]  
  \includegraphics[width=0.48\textwidth]{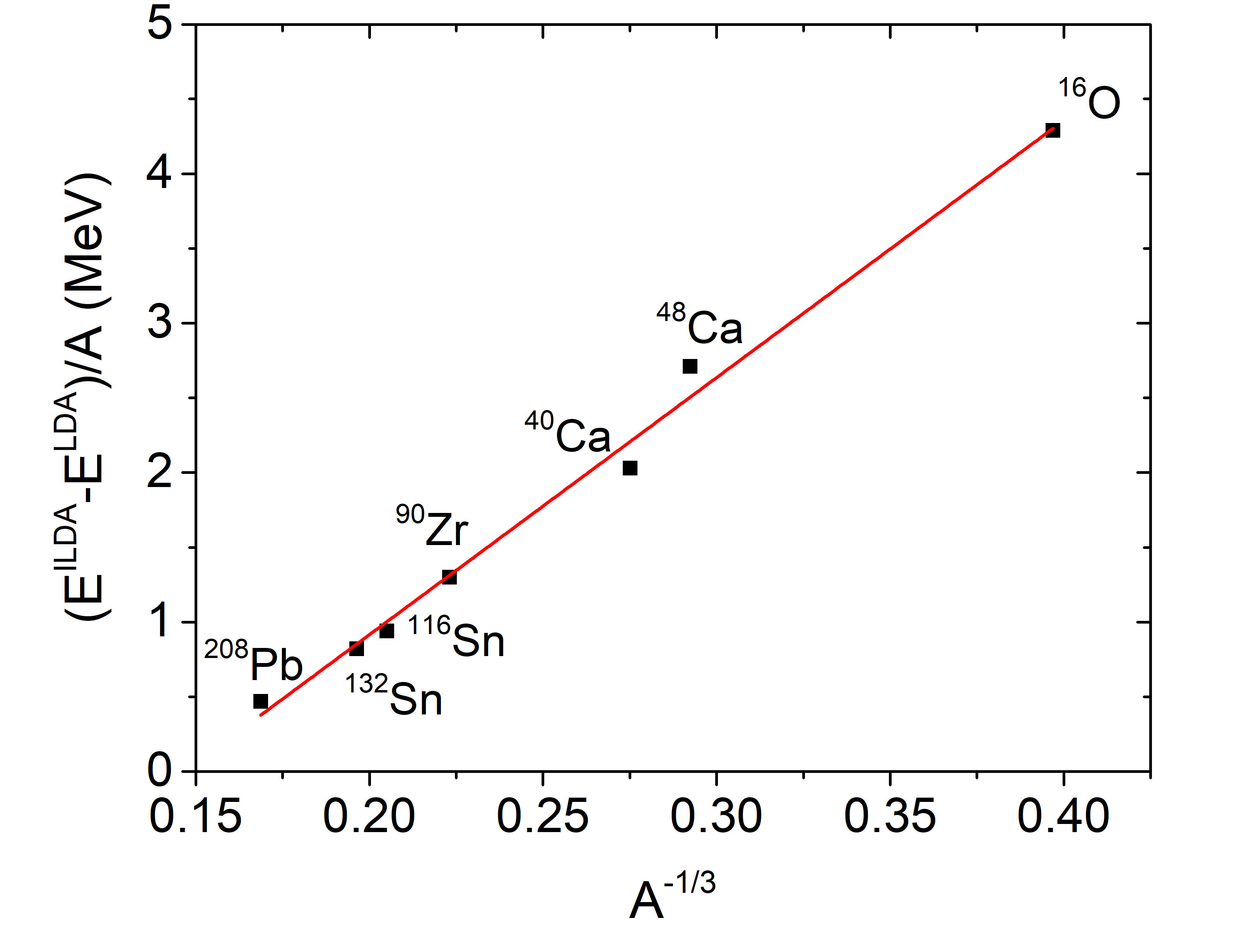}
  \caption{The deviations of the binding energy per nucleon in finite nuclei calculated in the ILDA and the LDA as a function of $A^{-1/3}$, where $A$ is the nuclear mass number. The inset shows the specific contributions of liquid-drop mass formula.}\label{Esurf}
\end{figure*}

The difference of the binding energy per nucleon for a given nucleus between the calculations in LDA and ILDA is plotted as a function of $A^{-1/3}$ in Fig. \ref{Esurf}.
It is found that introducing the ILDA, i.e. the finite range effect of interaction, reduces the binding energy especially for the light mass nuclei. The red line in Fig. \ref{Esurf} shows an obviously linear relationship, which can be described as
\begin{equation}\label{eqEsurf}
(E^\text{ILDA}-E^\text{LDA} )/A =  17.2\times A^{-1/3} - 2.5 ~~~ [\text{MeV}].
\end{equation}
Within the Bethe-Weizs\"acker liquid-drop mass formula, the surface energy ($B_\text{surf}=a_\text{surf} A^{2/3}$, $a_\text{surf}\approx17-20$ MeV) reduces the binding energy of nuclei. Thus the contribution of the surface energy to the binding energy per nucleon is proportional to $A^{-1/3}$. The constant of -2.5 MeV is required to compensate the effects of what we called ``surface density'' effect, which is included already in the LDA as discussed above.

We find that the values for the width parameter $t$ resulting from the fits described above are well described by
\begin{equation}\label{eqpret}
t=1.35-0.13A^{1/3}-0.17\frac{N-Z}{A}.
\end{equation}
Taking $^{88}$Sr as example, this eq.(\ref{eqpret}) yields a value of $t=0.75$ fm. Using this value for the ILDA description of this nucleus, one obtains a binding energy per nucleon $E^\text{ILDA}/A =-8.72$ MeV which is close to the experimental value $-8.73$ MeV. This indicates that the ILDA supplemented by eq.(\ref{eqpret}) could be a very simple tool to evaluate the properties of nuclei. As it is based on a microscopic many-body theory, it may even provide a reliable prediction for the binding properties of nuclei outside the stability valley.

\subsection{DISCUSSIONS ON $^{16}$O}
\begin{figure}[t]
  \centering
  \includegraphics[width=0.48\textwidth]{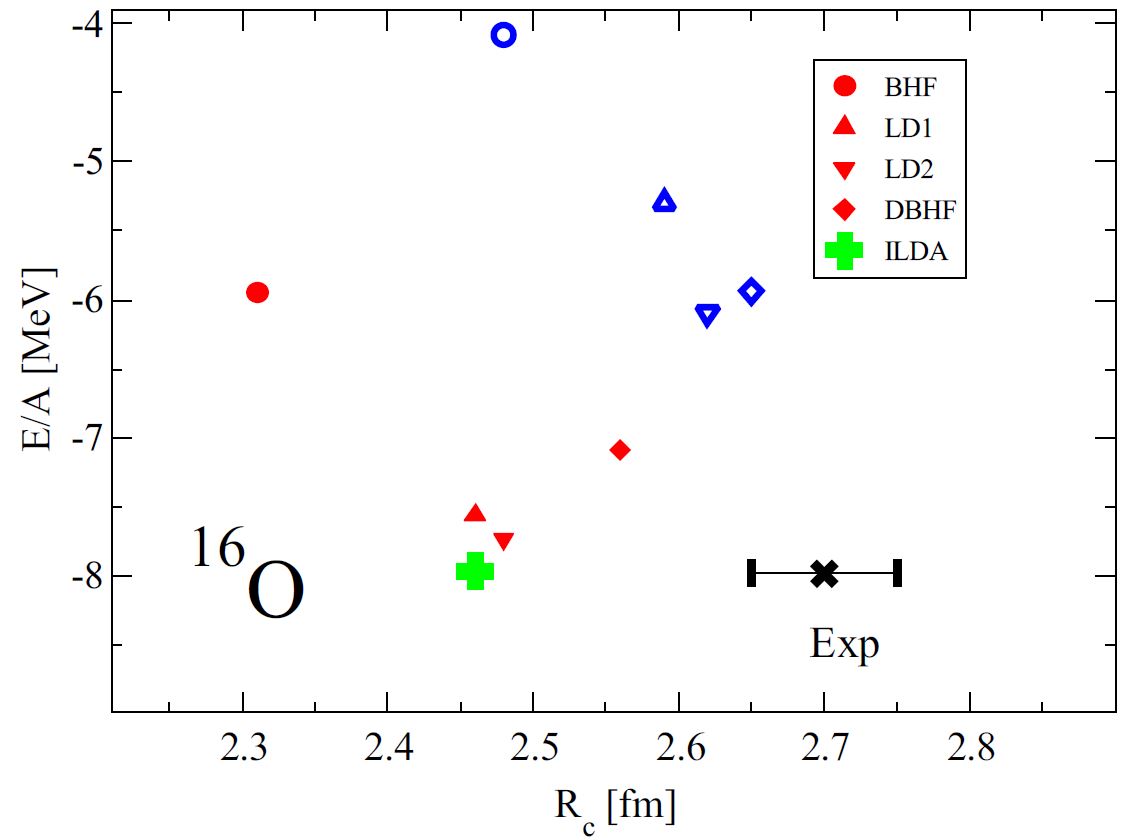}
  \caption{Binding energy per nucleon versus radius of charge distribution calculated for $^{16}$O by various methods based on OBEP model of the NN interaction. Further details see discussion in the text.}\label{o16}
\end{figure}

Finally, we compare our results for $^{16}$O with the results obtained by different approaches, which are all based on the OBE potentials $A$ and $C$ defined by Machleidt and Brockmann\cite{Machxxx,rolf84}. Results for the bulk properties, energy per nucleon ($E/A$) and radius of the charge distribution ($R_C$), are displayed in Fig. \ref{o16}
and compared to the experimental datum. Results for OBE potential $A$ are shown by filled symbols, whereas the corresponding results for the potential $C$ are represented by the open symbols. Results for conventional BHF calculations, which ignore the change of the Dirac spinors for the nucleons in the nuclear medium are denoted by circles and have been taken from the work of Fritz and M\"uther\cite{fritz2}. The self-consistent DBHF results of Shen $et$ $al$. \cite{shen1} are represented by the diamond symbols. As it has been discussed before, the self-consistent treatment of the Dirac spinor in the nuclear medium yields a substantial improvement for the simultaneous description of energy and radius of $^{16}$O. This improvement can also be observed for the two local density approximations, denoted as LD1 (denoted by upward triangles) and LD2 (downward triangle) in Fig. \ref{o16}. These approximations have been introduced in Ref. \cite{fritz2} and refer to BHF calculations, which employ a density dependence of the Dirac spinors taken from nuclear matter (LD1), whereas the approach LD2 corresponds to a $\sigma - \omega$ model as discussed in the introduction of this paper. These two approaches yield results, which are very close to the ILDA developed in this work. These results are also in reasonable agreement with the DBHF results of Ref. \cite{shen1}. 

It should be kept in mind that the LD1 and LD2 approach are based on a rather simple analysis of the Dirac structure in nuclear matter \cite{rolf84}, which is not as reliable as the subtracted $T$-matrix approach used in this work. Because of these limitations, the elder approaches had to be restricted to symmetric nuclear matter. As a consequence also LD1 and LD2 were restricted to the study of light isospin symmetric nuclei. 

In table \ref{tabO16}, the total energy per nucleon, the charge radius, the single-particle energies for protons $\pi$ and neutrons $\nu$ and the spin-orbit splitting of protons in the $p_{3/2}$ and $p_{1/2}$ subshells for $^{16}$O, $\Delta E_{\pi p}^{ls}$, are listed. The results of the DBHF calculations and the present approach (ILDA) are compared to the experimental data. Comparing the single-particle energies one finds that the ILDA single-particle energies are more bound than those resulting from DBHF and the experimental data. This can be attributed to the larger binding energy (as compared to DBHF) and smaller radius. Comparing single-particle energies with experiment one should keep in mind that the calculations are of the BHF kind. More elaborate definitions of the nucleon self-energy, like the so-called renormalized BHF, or the inclusion of hole-hole ladders lead to less attractive single-particle energies and larger radii. 

The spin-orbit splittings calculated in DBHF and ILDA are about 15 \% smaller than the empirical value. Also in this case one should be aware that more elaborate models for the single-particle spectrum, which e.g. include the coupling of the states to core excitations, may enhance the calculated splitting. Comparing the two calculations we think that the simple ILDA approach, which is applicable to all nuclei, yields a good agreement with those from the more sophisticated $ab$ $initio$ DBHF approach.

\begin{table}
  \centering
  \caption{Comparison of binding energy $E$, charge radius, $r_C$, single particle energies for neutrons $E_{\nu i}$ and protons $E_{\pi i}$, spin-orbit splitting of $^{16}$O $\Delta E_{\pi p}^{ls}$ between calculations and experimental data.}
  \label{tabO16}
  \setlength{\tabcolsep}{6pt}
  \begin{tabular}{cccc}
    \hline
    \hline
&&&\\
	&	Expt&	DBHF\cite{shen1}&	ILDA\\
&&&\\    \hline &&&\\  
$E$ [MeV]			&-127.6&-113.5	&-127.4 \\
$r_C$ (fm)	&2.70	&2.56	&2.46   \\
$E_{\nu s1/2}$	[MeV]&-47	&-48.1	&-53.4  \\
$E_{\nu p3/2}$ [MeV]	&-21.8	&-26.4	&-29.7  \\
$E_{\nu p1/2}$ [MeV]	&-15.7	&-21.0	&-24.5  \\
$E_{\pi s1/2}$ [MeV]	&-44$\pm$7	&-43.9	&-48.2  \\
$E_{\pi p3/2}$ [MeV]	&-18.5	&-22.5	&-25.1  \\
$E_{\pi p1/2}$ [MeV]	&-12.1	&-17.1	&-20.1  \\
$\Delta E_{\pi p}^{ls}$ [MeV]&6.3	&5.4	&5.0    \\
 &&&\\     \hline
    \hline
  \end{tabular}
\end{table}

\section{Summary}
In this paper we propose a new and simple scheme to describe the bulk properties of finite nuclei. This approach is based on the relativistic structure of the nucleon self-energy calculated within the microscopic DBHF scheme for asymmetric nuclear matter using the realistic OBE potential $A$\cite{Machxxx}. The Dirac scalar, $U_S$, and vector components, $U_0$, of the self-energy are determined using the subtracted T-matrix approach as described in Ref. \cite{subtrt}. A parameterization of these components depending on density, single-particle energy of the bound state and isospin asymmetry is presented. Assuming a local density approximation, which corresponds to the one employed for the optical model CTOM at positive energies\cite{ruirui2}, one can easily evaluate the scalar and vector component of the nucleon self-energy for the bound states of finite nuclei. The corresponding Dirac equations are solved and iterated until a solution with a self-consistent treatment of single-particle energies, densities for protons and neutrons is obtained. The surface effect of finite nuclei is considered within the Improved Local Density Approximation (ILDA) by folding the potential terms with a Gaussian function with width parameter $t$. The effect of this folding reproduces the surface energy term in the Weizs\"sacker mass formula with good accuracy. 

Using the ILDA approach one reproduces the bulk properties of spherical nuclei ranging from $^{16}$O to $^{208}$Pb with good precision. The total energies are reproduced very well, while the calculated radii for the charge distribution are typically too small by around 10 \%. Also the spin-orbit splitting of the single-particle energies are too small by around 15 \% as compared to the empirical value. These discrepancies are in line with corresponding results of a direct solution of the DBHF for finite nuclei\cite{shen1}. It is possible that these discrepancies could be corrected by considering a more sophisticated approach for the definition of the self-energy.

In the present study, the ILDA has been applied to the evaluation of spherical nuclei. It can easily be extended to the study of deformed nuclei with open shells. 

\begin{acknowledgments}
This work has been supported by the IAEA Coordinated Research Project F41032 (Grant Nos. 20466), the National Natural Science Foundation of China (Grant Nos. U1432247, 11775013, 11305270, 11465005, 11875321, and U1630143), and the Science Challenge Project with No. TZ2018001, and the Key Laboratory fund key projects with No. 6142A080201, and the Deutsche Forschungsgemeinschaft (DFG) under contract no. Mu 705/10-2.
\end{acknowledgments}

%\end{CJK*}


\begin{thebibliography}{999}
\bibitem{argonne} R.B. Wiringa, R.A. Smith, and T.L. Ainsworth, Phys Rev. {\bf C29}, 1207 (1984).
\bibitem{erkelenz} K. Erkelenz, Phys. Rep. {\bf 13}, 191 (1974).
\bibitem{mhelster} R. Machleidt, K. Holinde, and Ch. Elster, Phys. Rep. {\bf 149}, 1 (1987).
\bibitem{Machxxx} R. Machleidt, Adv. Nucl. Phys. {\bf 19}, 189 (1989).
\bibitem{kaiserbw} N. Kaiser, R. Brockmann, and W. Weise, Nucl. Phys. {\bf A625}, 758 (1997).
\bibitem{epelbaum} E. Epelbaum, W. Gl\"ockle, and U.-G. Mei\ss ner, Nucl. Phys. {\bf A671}, 295 (2000).
\bibitem{machchi} R. Machleidt and D.R. Entem, Phys. Rep. {\bf 503}, 1 (2011).
\bibitem{we35} C.F. von Weizs\"acker, Z. Phys. {\bf 96}, 431 (1935).
\bibitem{bethe36} H.A. Bethe and R.F. Bacher Rev. Mod. Phys. {\bf 8}, 82 (1936).
\bibitem{coester1} F. Coester, S. Cohen, B.D. Day, and C.M. Vincent, Phys. Rev. {\bf C1}, 769, (1970).
\bibitem{coester2} H. M\"uther and A. Polls, Prog. Part. Nucl. Phys.{\bf 45}, 243(2000).
\bibitem{3nuc1} J. Carlson, V.R. Pandharipande, and R. Wiringa, Nucl. Phys. {\bf A401}, 59 (1983).
\bibitem{3nuc2} M. Baldo and L.S. Ferreira, Phys. Rev. {\bf C59} 682 (1999).
\bibitem{3nuc3} E.N.E. van Dalen, P. G\"ogelein, and H. M\"uther, Phys. Rev. {\bf C80}, 044312 (2009).
\bibitem{mike} M.R. Anastasio, L.S. Celenza, W.S. Pong, and C.M. Shakin, Phys. Rep. {\bf 100}, 327 (1983).
\bibitem{horow84} C.J. Horowitz and B.D. Serot, Phys. Lett. {\bf B 137}, 287 (1984); Nucl. Phys. {\bf A464}, 613 (1987).
\bibitem{rolf84} R. Brockmann and R. Machleidt, Phys.  Lett. {B 149}, 283 (1984);Phys.  Rev. {\bf  C42}, 1965 (1990).
\bibitem{malf87} B. ter Haar and R. Malfliet, Phys. Rep. {\bf 149}, 207 (1987).
\bibitem{weigel88} P. Poschenrieder and M.K. Weigel, Phys. Rev. {\bf C38}, 471 (1988).
\bibitem{lenske} F. de Jong and H. Lenske, Phys. Rev. {\bf C58}, 890 (1998).
\bibitem{erik04} E.N.E. van Dalen, C. Fuchs, and A. Faessler, Nucl. Phys. {\bf A744}, 227 (2004).
\bibitem{walecka} B.D. Serot and J.D. Walecka, Adv. Nucl. Phys. {\bf 16}, 1 (1986).
\bibitem{francesca}  H. M\"uther, F. Sammarruca, and Zhongyu Ma, Int. J. Mod. Phys. {\bf E 26}, 17301 (2017).
\bibitem{shen1} S.H. Shen, H.Z. Liang, J. Meng, P. Ring, and S.Q. Zhang, Phys. Rev. {\bf C 96}, 014316 (2017).
\bibitem{shen2} S.H. Shen, H.Z. Liang, J. Meng, P. Ring, and S.Q. Zhang,  Phys. Lett. {\bf B 781}, 227 (2018). 
\bibitem{brockm} H. M\"uther, R. Machleidt, and R. Brockmann, Phys. Rev. {\bf C 42}, 1981 (1990).
\bibitem{boersma} H.F. Boersma and R. Malfliet, Phys. Rev. {\bf C 49}, 233 (1994).
\bibitem{fritz1} R. Fritz, H. M\"uther, and R. Machleidt, Phys. Rev. Lett. {\bf 71}, 46 (1993).
\bibitem{fritz2} R. Fritz and H. M\"uther, Phys. Rev. {\bf C 49}, 633 (1994).
\bibitem{toki} H. Shen, Y. Sugahara, and H. Toki, Phys. Rev. {\bf C 55}, 1211 (1997).
\bibitem{zyma} Z.Y. Ma and L. Liu, Phys. Rev. {\bf C 66}, 024321 (2002).
\bibitem{ruirui1}  R. R. Xu, Z. Y. Ma, E.N.E. van Dalen and H. M\"uther, Phys. Rev. {\bf C85}, 034613 (2012).
\bibitem{ruirui2} R. Xu, Z. Ma, Y. Zhang, Y. Tian, E.N.E. van Dalen and H. M\"uther, Phys. Rev. {\bf C94}, 034606 (2016).
\bibitem{Jeu77} J. P. Jeukenne, A. Lejeune, and C. Mahaux, Phys. Rev. {\bf C 16}, 80 (1977).
\bibitem{ctomtool} http://www.nuclear.csdb.cn/ctom/
\bibitem{subtrt} E.N.E. van Dalen and H. M\"uther, Phys. Rev. {\bf C82}, 014319 (2010).
\bibitem{Gross99} T. Gross-Boelting, C. Fuchs, Amand Faessler, Nucl. Phys. {\bf A648}, 105 (1999).
\bibitem{pair1}  O.A. Rubtsova, V.I. Kukulin, V. N. Pomerantsev, and H. M\"uther, Phys. Rev. {\bf C96}, 034327 (2017). 
\bibitem{dalen2011} E.N.E. van Dalen and H. M\"uther, Phys. Rev. {\bf C84}, 024320 (2011).
\bibitem{jamin82} M. Jaminon, Phys. Rev. {\bf C26}, 1551 (1982). 

\end{thebibliography}
\end{document}